\documentstyle[amstex,amssymb,12pt,psfig]{article}
\input{tcilatex}
\begin{document}
\date{}
\title{{\bf Cooper Pairing Revisited}}
\author{V.C. Aguilera-Navarro,$^{a}$ M. Fortes,$^{b}$ M. de Llano$^{c}$ \and
F.J. Sevilla$^{b}$   \\
$^{a}$Depto. de F\'{\i}sica, UNICENTRO, 85015, Guarapuava, PR, Brazil\\
$^{b}$Instituto de F\'{\i}sica, UNAM, 01000 M\'{e}xico, DF, Mexico\\
$^{c}$Instituto de Investigaciones en Materiales, UNAM,\\
04510 M\'{e}xico, DF, Mexico}
\maketitle

\section{Introduction}

Doubtlessly the most central notion in {\it superconductivity}, for both low
and high transition temperatures $T_{c}$, is that of Cooper pairs (CPs) \cite%
{Coo}\ that form among the underlying electron (or hole) charge carriers of
the many-electron system. And yet, it is perhaps the least understood
concept. Shortly after the publication of the BCS theory \cite{BCS}\ of
superconductivity, charged pairs observed in magnetic flux quantization
experiments with 3D conventional \cite{classical}\cite{classical2}, and much
later with quasi-2D cuprate \cite{cuprates} superconductors, suggested CPs
as an indispensable ingredient, regardless of the inability of BCS theory 
{\it per se }to describe high-$T_{c}$ superconductivity, as now seems an
almost universal consensus. Cooper pairing is no less central, albeit with a
different interfermion interaction, to {\it neutral-atom superfluidity} as
in liquid $^{3}He$\ \cite{He3}, and presumably also to ultracold {\it %
trapped alkali Fermi gases} such as $^{6}Li$ \cite{Li6}\ and $^{40}K$ \cite%
{Holland} where pairing is expected to occur as well. More recently, studies
of quantum degenerate Fermi gases consisting of neutral $^{40}K$\ atoms and
their so-called Feshbach ``resonance superfluidity'' have appeared \cite%
{Holland01}-\cite{Stringari02}, but assuming quadratically-dispersive CPs
and ignoring hole pairs---two fundamental shortcomings as we now illustrate.

In Sec. 2 we recall how the Bethe-Salpeter (BS) many-body equation (in the
ladder approximation),\ treating both 2p and 2h pairs on an equal footing,
implies that the ordinary CP problem [based on an ideal Fermi gas (IFG)
ground state (the usual ``Fermi sea'')] does {\it not} possess stable energy
solutions; in Sec. 3 we sketch how CPs based not on the IFG-sea but on the
BCS ground state survive, along with the usual trivial sound mode, as
nontrivial ``generalized'' or ``moving''\ CPs, linear in total or
center-of-mass-momentum (CMM) in leading order, that are {\it positive}
energy resonances with an imaginary energy term implying {\it %
finite-lifetime }effects. The nontrivial ``moving CP'' solution, though
often confused \cite{Randeria96}\ with it, is physically {\it distinct }from
the trivial sound mode solution sometimes called the
Anderson-Bogoliubov-Higgs (ABH) \cite{ABH}, (\cite{bts} p. 44), \cite{Higgs}%
\cite{Ran} collective excitation mode. The ABH mode is also linear in CMM in
leading order, and reduces to the IFG ordinary sound mode in zero coupling.
All this occurs in both the 3D study outlined in Ref. \cite{Honolulu} as
well as in 2D \cite{ANFdeLl}. Sec. 4 offers conclusions.

Provided CPs are bosons, as they indeed \cite{Clem} turn out to be, our
results will in general be crucial for Bose-Einstein condensation (BEC)
scenarios employing BF models of superconductivity, not only {\it in} {\it %
exactly 2D} as with the Berezinskii-Kosterlitz-Thouless \cite{BKT}\cite{KT}
transition, but also down to ($1+\epsilon $)D which characterize the
quasi-1D organo-metallic (Bechgaard salt) superconductors \cite%
{organometallics}-\cite{jerome2}. This is contrary to well-entrenched
perceptions (see, e.g., Ref. \cite{AAA}) that BEC is impossible in 2D.

\section{Ordinary Cooper pairing}

The {\it original} ``ordinary'' CP problem \cite{Coo}\ (with the BCS model
two-electron interaction) is defined in an $N$-electron system for two
electrons above the Fermi energy $E_{F}$\ but within a thin shell of energy $%
\hbar \omega _{D}.$ Only there do they suffer a constant attraction $-V<0$,
giving rise to the familiar result for the {\it negative}\ energy of the
pair, relative to the energy $2E_{F}$ of the interactionless pair, 
\begin{equation}
{\cal E}_{0}=-{\frac{2\hbar \omega _{D}}{e^{2/\lambda }-1}}\;\;%
\mathrel{\mathop{\longrightarrow}\limits_{\lambda \rightarrow 0}}\;\;-2\hbar
\omega _{D}e^{-2/\lambda }.  \label{eqn:Delta0}
\end{equation}%
Here $\lambda \equiv g(E_{F})V$ is a dimensionless coupling constant with $%
g(E_{F})$ the density of fermionic states (for each spin) evaluated at $%
E_{F} $. The equality in (\ref{eqn:Delta0}) is {\it exact} in 2D for all
coupling---as well as in 1D or 3D provided only that $\hbar \omega _{D}$ $%
\ll E_{F}$ so that $g(\epsilon )\simeq g(E_{F})$, a constant that can be
taken outside an energy integral.

However, the original CP problem\ neglects the effect of two-hole (2h) CPs
treated {\it simultaneously}\ on an equal footing with two-electron, or
two-particle (2p), CPs---as Green's functions \cite{FW}\ can naturally
guarantee. On the other hand, the BCS condensate consists of {\it equal
numbers of 2p and 2h Cooper correlations}. This was already evident, though
scarcely emphasized, from the perfect symmetry\ about $\mu $, the electron
chemical potential, of the well-known Bogoliubov \cite{Bog} $v^{2}(\epsilon )
$\ and $u^{2}(\epsilon )$ coefficients [see just below (\ref{19}) later on],
where $\epsilon $ is the electron energy.\ Indeed, our prime motivation
comes from the fact established recently \cite{Tolma} that the BCS
condensate is a BEC condensate for equal numbers of 2p and 2h pairs, in the
limit of weak coupling.\ Additional empirical motivation comes from the
unique but unexplained role played by {\it hole }charge carriers in the
normal state of superconductors in general \cite{Hirsch}. Even further
motivation stems from the ability of the ``complete (in that both 2h- and
2p-CPs are allowed in varying proportions) BF model'' of Refs. \cite{Tolma},%
\cite{PLA2}-\cite{CMT02} to ``unify'' both BCS and BEC theories as special
cases, and to predict substantially higher $T_{c}$'s than BCS theory without
abandoning electron-phonon dynamics. Compelling evidence for a significant
presence of this dynamics in high-$T_{c}$ cuprate superconductors from
angle-resolved photoemission spectroscopy data has recently been reported %
\cite{Shen}. 

In dealing with the many-electron system we assume the BCS model interaction
in the form with double Fourier transform 
\begin{equation}
\nu (\left| {\bf k}_{1}-{\bf k}_{1}^{\prime }\right|
)=-(k_{F}^{2}/k_{1}k_{1}^{\prime })V\text{ \ \ if}\;\ k_{F}-k_{D}<k_{1},%
\text{ }k_{1}^{\prime }<k_{F}+k_{D},\   \label{int}
\end{equation}%
and $=0\ $otherwise. As before $V>0$, $\hbar k_{F}\equiv mv_{F}$ the Fermi
momentum, $m$ the effective electron mass, $v_{F}$ the Fermi velocity, and $%
k_{D}\equiv \omega _{D}/v_{F}$ with $\omega _{D}$ the Debye frequency. The
usual physical constraint $\hbar \omega _{D}\ll E_{F}\equiv $ $\hbar
^{2}k_{F}^{2}/2m$ then implies that $k_{D}/k_{F}\equiv \hbar \omega
_{D}/2E_{F}\ll 1$.

The bound-state BS wavefunction equation \cite{Honolulu} in the ladder
approximation\ with both particles and holes for the original IFG-based CP
problem using an interaction such as (\ref{int})\ is%
\begin{gather}
\Psi ({\bf k,}E)=-\left( \frac{i}{\hbar }\right) ^{2}G_{0}\left( {\bf K}/2+%
{\bf k},{\cal E}_{K}/2+E\right) G_{0}\left( {\bf K}/2-{\bf k},{\cal E}%
_{K}/2-E\right) \times  \nonumber \\
\times \frac{1}{2\pi i}\dint\limits_{-\infty }^{+\infty }dE^{^{\prime }}%
\frac{1}{L^{d}}\sum_{{\bf k}^{\prime }}v(\left\vert {\bf k-k}^{\prime
}\right\vert )\Psi ({\bf k}^{\prime },E^{\prime })\,  \label{BSE}
\end{gather}%
where $L^{d}$ is the \textquotedblleft volume\textquotedblright\ of the $d$%
-dimensional system; ${\bf K\equiv k}_{1}+{\bf k}_{2}$ is the CMM and ${\bf %
k\equiv 
{\frac12}%
(k}_{1}-{\bf k}_{2}{\bf )}$ the relative wavevectors of the 2e bound state
whose wavefunction is $\Psi ({\bf k,}E)$; ${\cal E}_{K}$ $\equiv E_{1}+E_{2}$
is the energy of this bound state while $E\equiv E_{1}-E_{2}$, and $%
G_{0}\left( {\bf K}/2+{\bf k},{\cal E}/2+E\right) $ is the bare one-fermion
Green's function given by%
\begin{equation}
G_{0}({\bf k}_{1},E_{1})=\frac{\hbar }{i}\left\{ \frac{\theta (k_{1}-k_{F})}{%
-E_{1}+\epsilon _{{\bf k}_{1}}-E_{F}-i\varepsilon }+\frac{\theta
(k_{F}-k_{1})}{-E_{1}+\epsilon _{{\bf k}_{1}}-E_{F}+i\varepsilon }\right\}
\label{IFGgreen}
\end{equation}%
where $\epsilon _{{\bf k}_{1}}\equiv $ $\hbar ^{2}k_{1}^{2}/2m$ and $\theta
(x)=1$ for $x>0$ and $=0$ for $x<0$, so that the first term refers to {\it %
electrons} and the second to {\it holes}. Figure 1 shows all Feynman
diagrams for the 2p, 2h and ph wavefunction $\psi _{+}$, $\psi _{-}$ and $%
\psi _{0}$, respectively, that emerge in the general
(BCS-ground-state-based) problem to be discussed later. For the present
IFG-based case, diagrams in shaded rectangles do {\it not} contribute. Since
the energy dependence of $\Psi ({\bf k},E)$ in (\ref{BSE}) is only through
the Green's functions, the ensuing energy integrals can be evaluated
directly in the complex $E^{\prime }$-plane and yield, for interaction (\ref%
{int}),\ 
\begin{equation}
(2\xi _{k}-{\cal E}_{0})\psi _{{\bf k}}=V\sum_{{\bf k}^{\prime
}}{}^{^{\prime }}\psi _{{\bf k}^{\prime }}-V\sum_{{\bf k}^{\prime
}}{}^{^{\prime \prime }}\psi _{{\bf k}^{\prime }}  \label{CP}
\end{equation}%
where $\,\psi _{{\bf k}}$ is the resulting wavefunction after the energy
integration. Here $\xi _{k}\equiv \hbar ^{2}k^{2}/2m-E_{F}$ while ${\cal E}%
_{0}$ is the (unknown) eigenvalue energy. The single prime over the first
(2p-CP) summation term denotes the restriction $0<\xi _{k^{\prime }}<\hbar
\omega _{D}$ (i.e., {\it above }the IFG \textquotedblleft
sea\textquotedblright ) while the double prime in the last (2h-CP) term
means $-\hbar \omega _{D}<\xi _{k^{\prime }}<0$ (i.e., {\it below }the IFG
sea). Without this latter term we have Cooper's Schr\"{o}dinger-like
equation \cite{Coo}\ for 2p-CPs whose implicit wavefuncion solution is
clearly $\psi _{{\bf k}}=(2\xi _{k}-{\cal E}_{0})^{-1}V\sum_{{\bf k}^{\prime
}}^{^{\prime }}\psi _{{\bf k}^{\prime }}.$ Since the summation term is
constant, performing that summation on {\it both} sides allows canceling the 
$\psi _{{\bf k}}$-dependent terms, leaving the eigenvalue equation $\sum_{%
{\bf k}}^{^{\prime }}(2\xi _{k}-{\cal E}_{0})^{-1}=1/V$. This is one
equation in one unknown ${\cal E}_{0}$; transforming the sum to an integral
over energies immediately gives (\ref{eqn:Delta0}). This corresponds to the
usual negative-energy, infinite-lifetime stationary-state bound pair. For $%
K\geqslant 0$ the CP eigenvalue equation is just $\sum_{{\bf k}}{}^{^{\prime
}}(2\xi _{k}-{\cal E}_{K}+\hbar ^{2}K^{2}/4m)^{-1}=1/V.$ Since a CP state of
energy ${\cal E}_{K}$\ is characterized only by a definite $K$ but {\it not }%
definite ${\bf k}$, in contrast to a \textquotedblleft BCS
pair\textquotedblright\ defined [Ref. \cite{BCS}, Eqs. (2.11) to (2.13)]\
with fixed\ ${\bf K}$ and ${\bf k}$ (or equivalently definite ${\bf k}_{1}$\
and ${\bf k}_{2}$). This renders CPs legitimate bosons \cite{Clem}. Without
the first summation term in (\ref{CP})\ the same result in ${\cal E}_{0}$\ (%
\ref{eqn:Delta0})\ for 2p-CPs follows for 2h-CPs (apart from a sign change).

In contrast with Cooper's equation neglecting hole-pairs, the {\it complete }%
CP{\it \ }equation (\ref{CP}) {\it cannot }be derived from an ordinary
(non-BS) Schr\"{o}dinger-like equation in spite of its simple appearance. A
more general technique such as the BS equation that includes both particles
(in this case electrons) and holes is needed. To solve it for the unknown
energy ${\cal E}_{0}$, let the rhs of (\ref{CP}) be defined as $A-B$, with $A
$ relating to the 2p-pair term and $B$ to the 2h-pair term. Then the unknown 
$\psi _{{\bf k}}$ becomes%
\begin{equation}
\psi _{{\bf k}}=(A-B)/(2\xi _{k}-{\cal E}_{0})\text{ \ \ \ or equivalently \
\ \ }\psi (\xi )=(A-B)/(2\xi -{\cal E}_{0})  \label{15}
\end{equation}%
whence 
\begin{eqnarray}
A &\equiv &\lambda \int_{0}^{\hbar \omega _{D}}d\xi \psi (\xi )=%
{\frac12}%
(A-B)\lambda \int_{-{\cal E}_{0}}^{2\hbar \omega _{D}-{\cal E}%
_{0}}dz/z\equiv (A-B)x,  \nonumber \\
B &\equiv &\lambda \int_{-\hbar \omega _{D}}^{0}d\xi \psi (\xi )=%
{\frac12}%
(A-B)\lambda \int_{-2\hbar \omega _{D}-{\cal E}_{0}}^{-{\cal E}%
_{0}}dz/z\equiv (A-B)y.  \label{17}
\end{eqnarray}%
The integrals are readily evaluated giving $x\equiv 
{\frac12}%
\lambda \ln (1-2\hbar \omega _{D}/{\cal E}_{0})$ and $y\equiv -%
{\frac12}%
\lambda \ln (1+2\hbar \omega _{D}/{\cal E}_{0})$. As $A$ and $B$ still
contain the unknown $\psi (\xi )$ let us eliminate them. Note that equations
(\ref{17}) are equivalent to {\it two} equations in two unknowns $A$ and $B$%
, or 
\[
(1-x)A+xB=0\text{ \ \ \ and \ \ \ }-yA+(1+y)B=0.
\]%
These readily lead to the single equation $1-x+y=0,$ which on inserting the
definitions for $x$ and $y$ becomes 
\begin{equation}
1=%
{\frac12}%
\lambda \ln [1-(2\hbar \omega _{D}/{\cal E}_{0})^{2}]\text{ \ \ \ which
gives \ \ \ }{\cal E}_{0}=\pm i2\hbar \omega _{D}/\sqrt{e^{2/\lambda }-1}.
\label{18}
\end{equation}%
As the CP energy is pure-imaginary, there is an obvious instability of the
CP problem when both particle- and hole-pairs are included. This
transcendant result dates back to the late 50's and early 60's and was
reported in Refs. \cite{bts} p. 44 and \cite{AGD} Sec. 33, where, however,
the well-known pure 2p and 2h special cases just stressed was not discussed.
Clearly then, the original CP picture {\it is meaningless if particle- and
hole-pairs are treated on an equal footing}, as consistency demands.
Curiously, this result has been largely ignored in the entire literature
since then.

A dramatic analogue of the nontriviality of such consistency is found in the
very high temperature treatment of relativistic BEC \cite{HaberWeldon},
where pair production becomes possible and creates {\it antibosons }in
addition to more bosons. Here, BEC {\it must }take into account$\ \overset{%
\_\_}{N}\ $antibosons of charge, say, $-q$ along with the $N$ bosons of
charge $q$. In units such that $\hbar \equiv c\equiv k_{B}\equiv 1$ the
boson energy is $\varepsilon _{K}=(K^{2}+m_{B}^{2})^{1/2}$. Charge
conservation requires that not only $N\equiv N_{0}(T)+\sum_{{\bf K\neq 0}%
}[\exp \beta (\varepsilon _{K}-\mu _{B})-1]^{-1}$ be constant but rather $N-$
$\overset{\_\_}{N\text{,}}$ where $\overset{\_\_}{N}$ is the same as $N$ but
with $+\mu _{B}$ instead of $-\mu _{B}$. If $\rho \equiv q(N-\overset{\_\_}{N%
})/L^{3}\equiv qn$ is the net conserved charge density, it is shown in Ref. %
\cite{HaberWeldon} that $T_{c}=(3|n|/m_{B})^{%
{\frac12}%
}$ and that the condensate fraction $n_{0}/n=[1-(T/T_{c})^{2}].$ This is 
{\it qualitatively} different from the better-known results assuming only $N$
constant, which are the mass-independent $T_{c}=[\pi ^{2}n/\zeta (3)]^{1/3}$
and $n_{0}/n=[1-(T/T_{c})^{3}].$ This example exhibits the strikingly
dramatic effect of including or not antiparticles (analogous to holes in the
nonrelativistic case).

\section{Generalized Cooper pairing}

However, a BS treatment not about the IFG sea but about the BCS ground state
(which we refer to as ``generalized''\ Cooper pairing) {\it vindicates the
CP concept}, and adds something new. This is equivalent to starting not from
the IFG unperturbed Hamiltonian but from the BCS one. Thus, (\ref{IFGgreen})
is replaced by 
\begin{equation}
\text{{\bf G}}_{0}({\bf k}_{1},E_{1})=\frac{\hbar }{i}\left\{ \frac{%
v_{k_{1}}^{2}}{-E_{1}+E_{k_{1}}-i\varepsilon }+\frac{u_{k_{1}}^{2}}{%
-E_{1}+E_{k_{1}}+i\varepsilon }\right\}   \label{BCSgreen}
\end{equation}%
where $E_{{\bf k}}\equiv \sqrt{\xi _{k}{}^{2}+\Delta ^{2}}$ with $\Delta $
the fermionic gap, $v_{k}^{2}\equiv 
{\frac12}%
(1-\xi _{k}/E_{{\bf k}})$ and $u_{k}^{2}\equiv 1-v_{k}^{2}$ are the
Bogoliubov functions \cite{Bog}.\ As $\Delta \rightarrow 0$\ these three
quantities become $|\xi _{k}|$, $\theta (k_{1}-k_{F})$ and $\theta
(k_{F}-k_{1})$, respectively, making (\ref{BCSgreen}) become (\ref{IFGgreen}%
)\ as expected. Substituting $G_{0}({\bf k}_{1},E_{1})$ by {\bf G}$_{0}({\bf %
k}_{1},E_{1})$\ corresponds to rewriting the total Hamiltonian so that the
pure-kinetic-energy unperturbed Hamiltonian is replaced by the BCS one. The
remaining terms are then assumed suitable to a perturbation treatment.
Experimental support for this can be found precisely in Refs. \cite%
{classical}-\cite{cuprates}, and its physical justification lies in
recovering both the expected ABH sound mode (which contains the BCS $T=0$
gap equation)\ and the finite-lifetime effects of moving CPs. In either 3D %
\cite{Honolulu}\ or 2D \cite{ANFdeLl}\ the BCS-based BS equation yields {\it %
two} {\it distinct} {\it solutions}: a) the usual trivial ABH sound solution
and b) a highly nontrivial ``moving CP''\ solution. In either case the BS
formalism consists of a set of three coupled equations, one for each (2p, 2h
and ph) channel\ wavefunction\ for any spin-independent interaction such as (%
\ref{int}). However, the ph channel decouples, leaving only two coupled
wavefunction\ equations for the ABH solution in 2D which we consider first.
We focus here on 2D because of its interest \cite{Brandow}\ for quasi-2D
cuprate superconductors.

The equations involved are too lengthy even in 2D\ and will be derived in
detail elsewhere, but for the trivial or ABH sound solution they boil down
to the single expression 
\begin{gather}
\frac{1}{2\pi }\lambda \hbar
v_{F}\int_{k_{F}-k_{D}}^{k_{F}+k_{D}}dk\int_{0}^{2\pi }d\varphi \{u_{{\bf K}%
/2+{\bf k}}{\Huge \,}u_{{\bf K}/2-{\bf k}}\text{ }+\text{ }v_{{\bf K}/2+{\bf %
k}}{\Huge \,}v_{{\bf K}/2-{\bf k}}\}\times   \nonumber \\
\times \left[ \frac{v_{{\bf K}/2+{\bf k}}v_{{\bf K}/2-{\bf k}}}{{\cal E}%
_{K}+E_{{\bf K}/2+{\bf k}}\text{ }+\text{ }E_{{\bf K}/2-{\bf k}}}+\,\frac{u_{%
{\bf K}/2+{\bf k}}\,u_{{\bf K}/2-{\bf k}}}{-{\cal E}_{K}+E_{{\bf K}/2+{\bf k}%
}+E_{{\bf K}/2-{\bf k}}}\right] =1  \label{19}
\end{gather}%
where $\varphi $ is the angle between ${\bf K}$ and ${\bf k}$. Here $k_{D}$
is defined as below (\ref{int}) and as before $\lambda \equiv Vg(E_{F})$
with $g(E_{F})\equiv m/2\pi \hbar ^{2}$ the constant 2D electronic DOS and $V
$ is defined in (\ref{int}). The {\it ABH collective excitation mode }energy 
${\cal E}_{K}$\ must then be extracted from this equation. For ${\bf K}=0$
it is just ${\cal E}_{0}=0$ (Ref. \cite{bts} p. 39) and (\ref{19}) rewritten
as an integral over $\xi \equiv \hbar ^{2}k^{2}/2m-E_{F}$ reduces to the
familiar BCS $T=0$ gap equation $\int_{0}^{\hbar \omega _{D}}d\xi /\sqrt{\xi
^{2}+\Delta ^{2}}=1/\lambda $\ for interaction (\ref{int}) which gives\ $%
\Delta =\hbar \omega _{D}/\sinh (1/\lambda ).$ Returning to the ABH energy $%
{\cal E}_{K}$ equation (\ref{19})\ and Taylor-expanding ${\cal E}_{K}$ about 
$K=0,$ and then taking $\Delta $ small$,$ leaves 
\begin{equation}
{\cal E}_{K}=\frac{\hbar v_{F}}{\sqrt{2}}K+O(K^{2})+o(\lambda )  \label{22}
\end{equation}%
where\ $o(\lambda )$ refers to interfermion interaction terms that vanish as 
$\lambda \rightarrow 0.$ Note that the leading term is just the ordinary
sound mode in an IFG whose sound speed $c$ $=$ $v_{F}/\sqrt{d}$ in $d$
dimensions. This result also follows (trivially) on solving for $c$ in the
familiar thermodynamic relation $dP/dn=mc^{2}$ involving the
zero-temperature IFG pressure $P=n^{2}[d(E/N)/dn]=2nE_{F}/(d+2)=2C_{d}n^{2/d%
\text{ }+1}/(d+2)$ where the constant $C_{d}$ will drop out$.$ Here the IFG
ground-state energy $E=dNE_{F}/(d+2)$ was used along with $E_{F}\equiv \hbar
^{2}k_{F}^{2}/2m=C_{d}n^{2/d}$ while $n\equiv N/L^{d}=k_{F}^{d}/d2^{d-2}\pi
^{d/2}\Gamma (d/2)$ is the fermion-number density. The derivative $dP/dn$
finally gives $c$ $=$ $\hbar k_{F}/m\sqrt{d}\ \equiv v_{F}/\sqrt{d}$ which
in 2D is just the coefficient of the first term of (\ref{22}).

The nontrivial or {\it moving CP}\ solution of the BCS-ground-state-based BS
treatment, which is {\it entirely new}, leads to the pair energy ${\cal E}%
_{K}$ which in 2D is contained in the equation 
\begin{gather}
\frac{1}{2\pi }\lambda \hbar
v_{F}\int_{k_{F}-k_{D}}^{k_{F}+k_{D}}dk\int_{0}^{2\pi }d\varphi u_{{\bf K}/2+%
{\bf k}}v_{{\bf K}/2-{\bf k}}\times   \nonumber \\
\times \{u_{{\bf K}/2-{\bf k}}v_{{\bf K}/2+{\bf k}}-u_{{\bf K}/2+{\bf k}}v_{%
{\bf K}/2-{\bf k}}\}\frac{E_{{\bf K}/2+{\bf k}}+E_{{\bf K}/2-{\bf k}}}{-%
{\cal E}_{K}^{2}+(E_{{\bf K}/2+{\bf k}}+E_{{\bf K}/2-{\bf k}})^{2}}=1,
\label{mCP}
\end{gather}%
In addition to the pp and hh wavefunctions (depicted diagrammatically in
Ref. \cite{Honolulu} Fig. 2), diagrams associated with the ph channel give
zero contribution at $T=0$. A third equation for the ph wavefunction
describes the ph bound state but turns out to depend only on the pp and hh
wavefunctions. Taylor-expanding ${\cal E}_{K}$ in (\ref{mCP}) in powers of $K
$\ around $K=0$, and introducing a possible damping factor by adding an
imaginary term $-i\Gamma _{K}$ in the denominator, yields to order $K^{2}$
for small $\lambda $%
\begin{equation}
\pm {\cal E}_{K}\simeq 2\Delta +\frac{\lambda }{2\pi }\hbar v_{F}K+\frac{1}{9%
}\frac{\hbar v_{F}}{k_{D}}e^{1/\lambda }K^{2}-i\hbar v_{F}K\left[ \frac{%
\lambda }{\pi }+\frac{1}{12k_{D}}e^{1/\lambda }K\right] +O(K^{3})
\label{linquadmCP}
\end{equation}%
where the upper and lower sign refers to 2p- and 2h-CPs, respectively. A
linear dispersion in leading order again appears, but now associated with
the bosonic moving CP. Figure 2a graphs\ the exact moving CP {\it real}
energy\ (full curves) extracted from (\ref{mCP}), along with its leading
linear-dispersion term (short-dashed) and this plus the next (quadratic)
term (long-dashed) from (\ref{linquadmCP}). The interaction parameter values
used for (\ref{int})\ were $\hbar \omega _{D}/E_{F}=0.05$ (a typical value
for cuprates) and the two values $\lambda =%
{\frac14}%
$ (lower set of curves) and $%
{\frac12}%
$ (upper set), so that ${\cal E}_{0}/E_{F}\equiv 2\Delta /E_{F}=2\hbar
\omega _{D}/E_{F}\sinh (1/\lambda ).$This has values $\simeq 0.004$ (for $%
\lambda =%
{\frac14}%
$) and $0.028$ (for $\lambda =%
{\frac12}%
$)$,$ (marked as dots in the figure). Remarkably enough, the linear
approximation (short-dashed lines in figure) is better over a wider range of 
$K/k_{F}$ values for weaker coupling in spite of a larger and larger partial
contribution from the quadratic term in (\ref{linquadmCP}). This peculiarity
also emerged from the ordinary CP treatment with the BCS model interaction %
\cite{PC98} in both 2D and 3D, and with a (regularized) delta interaction in
2D \cite{PRB2000} and in 3D \cite{PhysicaC}, and might suggest the expansion
in powers of $K$ to be an asymptotic series that should be truncated after
the linear term. For reference we also plot the linear term $\hbar v_{F}K/%
\sqrt{2}$\ of the sound solution (\ref{22}) (thick long-dashed line starting
at origin). The {\it positive}-energy 2p-CP resonance of width $\Gamma _{K}$%
\ has a lifetime from (\ref{linquadmCP}) of $\tau _{K}\equiv \hbar /2\Gamma
_{K}=\hbar /2\left[ (\lambda /\pi )\hbar v_{F}K+(\hbar
v_{F}/12k_{D})e^{1/\lambda }K^{2}\right] $ diverging only at $K=0$, and
falling to zero as $K$ increases; see Fig. 2b. Thus, ``faster''\ moving CPs
are shorter-lived and eventually break up, while ``non-moving''\ ones are
(infinite-lifetime) stationary states. The real linear term $(\lambda /2\pi
)\hbar v_{F}K$ in (\ref{linquadmCP})\ contrasts sharply with the {\it %
coupling-independent} leading-term $(2/\pi )\hbar v_{F}K$\ that follows \cite%
{PC98} from the {\it original }CP problem neglecting holes, which if graphed
in Fig. 2a would almost coincide with the ABH term $\hbar v_{F}K/\sqrt{2}$
and have a slope about $90\%$ smaller. Fig. 2c depicts analogies of a 3D
potential problem for the original CP energy (top), as well as for $K>0$
(middle) and $K=0$ (bottom) BCS-based BS CPs.

As in Cooper's \cite{Coo}\ original equation, our BCS-based BS moving CPs
are characterized by a definite ${\bf K}$ and {\it not} also by definite $%
{\bf k}$ as the pairs discussed by BCS \cite{BCS}. Hence, the objection does
not apply that CPs are not bosons because BCS pairs with definite\ ${\bf K}$
and ${\bf k}$ (or equivalently definite ${\bf k}_{1}$\ and ${\bf k}_{2}$)
have creation/annihilation operators that do {\it not }obey the usual Bose
commutation relations [Ref. \cite{BCS}, Eqs. (2.11) to (2.13)]. In fact, (%
\ref{mCP}) shows that a given ``generalized'' CP state labeled by either $%
{\bf K}$ or ${\cal E}_{K}$\ can accommodate (in the thermodynamic limit) an
indefinitely many possible BCS pairs with different ${\bf k}$'s; see Ref. %
\cite{Clem}. A recent electronic analog \cite{Samuelsson} of the Hanbury
Brown-Twiss photon-effect experiment suggests electron pairs to be
definitely bosons.

\section{Conclusions}

Hole pairs treated on a par with electron pairs play a vital role in
determining the precise nature of nontrivial CPs even at zero
temperature---only when based not on the usual IFG ``sea''\ but on the BCS
ground state. Their treatment with a Bethe-Salpeter equation gives
purely-imaginary-energy CPs when based on the IFG, and when based on the BCS
ground state gives\ positive-energy, resonant-state CPs with a finite
lifetime for nonzero CMM. This is instead of the more familiar
negative-energy stationary states of the original IFG-based CP problem that
neglects holes, as sketched just below (\ref{CP}). The BS ``moving-CP''\
dispersion relation (\ref{linquadmCP}), on the other hand, is gapped by
twice the BCS energy gap, followed by a {\it linear} leading term in the CMM
expansion about $K=0$. This linearity is distinct from the better-known but
trivial one (\ref{22}) associated with the sound or ABH collective
excitation mode whose energy vanishes at $K=0$. 

Thus, instead of the quadratic\ $\hbar ^{2}K^{2}/2(2m)$ assumed for CPs in
Refs. \cite{Holland01}-\cite{Stringari02},\cite{Tolma},\cite{PLA2},\cite%
{Blatt}-\cite{BF6}, among many others,\ BF models based on the correct CP
linearity for the boson component can give BEC for all $d>1$, including\
exactly 2D, and thus in principle address not only quasi-2D cuprate but also
quasi-1D organo-metallic superconductors.

\bigskip

{\bf Acknowledgments }MdeLl\ and MF thank UNAM-DGAPA-PAPIIT (Mexico), grant
\# IN106401, and CONACyT (Mexico), grant \# 27828 E, for partial support.
MdeLl is grateful for travel support through a grant to Southern Illinois
University at Carbondale from the U.S. Army Research Office.

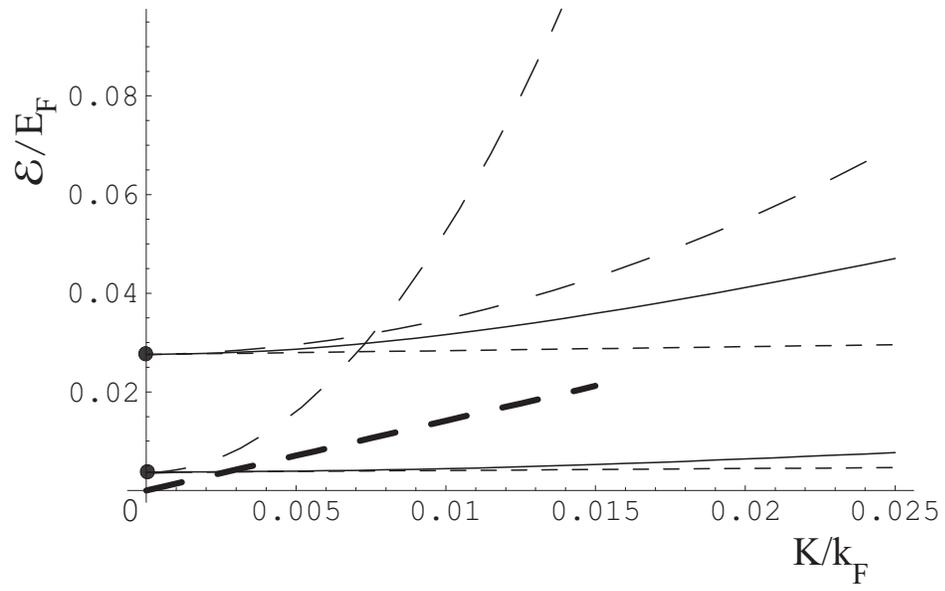
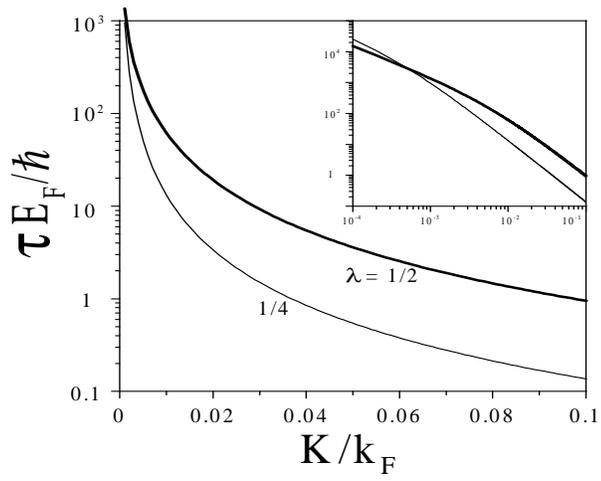
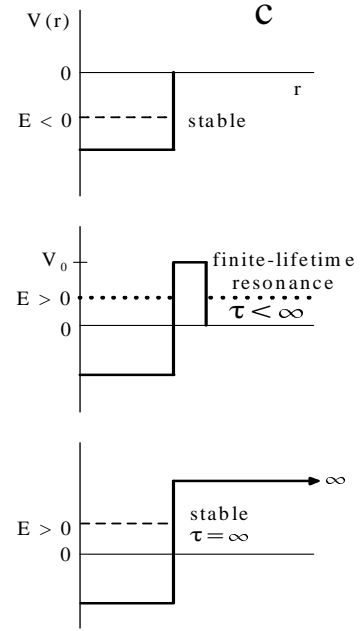
\end{document}